\documentclass[prc,twocolumn,secnumroman,tightenlines,amssymb,aps,nobibnotes,
superscriptaddress,showpacs,balancelastpage,floatfix]{revtex4}
\usepackage{graphicx}
\usepackage{multirow}           
\usepackage{color}	
\expandafter\ifx\csname package@font\endcsname\relax\else
\expandafter\expandafter
\expandafter\usepackage
\expandafter{\csname package@font\endcsname}
\fi
\DeclareRobustCommand\substyle{\name@idx{document substyle}}
\DeclareRobustCommand\classoption{\name@idx{document class option}}
\DeclareRobustCommand\classname{\name@idx{document class}}
\def\name@idx#1#2{{\ttfamily#2}
\index{#2\space#1=\string\ttt{#2}\space#1}\index{#1>#2=\string\ttt{#2}}}
\newcommand{\mc}[3]{\multicolumn{#1}{#2}{#3}}
\newcommand{\MatUnit}{1\kern-3pt 1} 
\newcommand{\Bra}[1]{\langle #1 \vert} 
\newcommand{\Ket}[1]{\vert #1 \rangle} 

\begin{document}

%
%
%
%
%

\title[]
      {Rotational bands in Quadrupole-Octupole Collective Model}
\author{A. Dobrowolski}
\email{arturd@kft.umcs.lublin.pl}
\affiliation{Department of Theoretical Physics, Maria Curie-Sk\l{}odowska University,\\ pl. Marii Curie-Sk\l{}odowskiej 1,
             PL-20031 Lublin, Poland}

\author{K. Mazurek}
\email{Katarzyna.Mazurek@IFJ.edu.pl}
\affiliation{Institute of Nuclear Physics PAN, ul. Radzikowskiego 152,
             Pl-31342 Krak\'ow, Poland}

\author{A. G\'o\'zd\'z}
\email{Andrzej.Gozdz@umcs.lublin.pl}
\affiliation{Department of Theoretical Physics, Maria Curie-Sk\l{}odowska University,\\ pl. Marii Curie-Sk\l{}odowskiej 1,
             PL-20031 Lublin, Poland}

\vspace{10pt}
\date{\today}

\begin{abstract}
A collective bands of positive and negative parity could be composed of the vibrations and rotations. The rotations of the octupole configurations can be based either on 
the axial or the non-axial octupole vibrations.
A consistent approach to the quadrupole-octupole collective vibrations coupled with the rotational motion enables to distinguish between various scenarios of disappearance of the E2 transitions in negative-parity bands. The here presented theoretical estimates are compared with the recent experimental energies and transition probabilities in and between the ground-state and low-energy negaive-parity bands in $^{156}$Dy.
A realistic collective Hamiltonian contains the potential energy term obtained through the macroscopic-microscopic Strutinsky-like method with particle-number-projected BCS approach and deformation-dependent mass tensor defined in vibrational-rotational, nine-dimensional collective space. The symmetrization procedure ensures the uniqueness of the Hamiltonian eigensolutions with respect to the laboratory coordinate system.
This quadrupole-octupole collective approach may also allow to find and/or verify some fingerprints of possible high-order symmetries (e.g. tetrahedral, octahedral,...) 
 in nuclear collective bands.

\end{abstract}


\maketitle


\section{Introduction}
\label{Section.01}

The idea of looking for the experimental evidence of the high-rank symmetries in atomic nuclei has been triggered by a series of theoretical articles i.e.: \cite{Roh97,dudap2003,JDS03,dudrl2002,dudek2007,pango2007}, where existence of the nonaxial octupole stable configurations have been connected with negative-parity bands.
The octupole deformation of the nucleus was confirmed experimentally by studies of experimental observables such as: rotational spectra of quadrupole or octupole deformed nuclei, \cite{Roh88,But96}, the transition probabilities, magnetic moments and some properties of K-isomers \cite{Min12,Bon15}. Recently, the investigation of negative-parity, odd-spin states in $^{156}$Dy has been repeated with the high resolving power of the Gammasphere array \cite{Har17}. 

The most expected evidence of the non-axial octupole deformation of rotating nucleus would be the disappearance of the E2 transitions between the lowest states in negative parity band. The E2 transitions are measured in ''band 2'' of Ref.~\cite{Har17} for $^{156}$Dy from state 27$^-$ down to 7$^-$ but below they are invisible. 

Since in our previous paper of Ref.\cite{dobrowolski:2016} the ground-state and negative-parity bands for $^{156}$Gd nucleus has been discussed in collective quadrupole-octupole model for low spins (0$<$J$<$5), we decided to extend our study to higher spins. We hope that we are able to answer the question posed by experimentalist in Ref.~\cite{Har17} in IVB section concerning ''band 2'': 'Octupole vibrations or tetrahedral symmetry'.  

Nuclear vibrations has been discussed by several authors in \cite{naz,nazar}, using the Bohr Hamiltonian \cite{Pro09,Bizzeti1,Bizzeti2} or by the Interacting Boson Model (IBM) \cite{Eng85,Eng87} or analytic collective model (AQOA) \cite{Bon05}. Also the new approach based on cluster Hamiltonian are shown in \cite{shneid15}. All those approaches are aimed for searching stable nuclear configurations and the strengths of electromagnetic transitions between collective states where quadrupole and octupole deformation parameters play the role of collective variables.	 

The collective Hamiltonian used in the present paper contains the collective potential obtained within macroscopic-microscopic total energy calculations using the Strutinsky method with the Woods-Saxon mean field \cite{cwiok87,JDPRL} and the cranking approximation  \cite{Inglis} for the mass tensors in kinematic part. The vibrational-rotational Hamiltonian are constructed in the intrinsic frame applying the so called 'adiabatic approximation'. The spherical harmonics parametrization of the nuclear deformation allow to control the geometrical properties of nuclear surface and applying the symmetrization procedure. The deformation space is limited to the dipole ($\lambda=1$), quadrupole ($\lambda=2$) and octupole ($\lambda=3$) parameters. 

The collective Hamiltonian is diagonalized in the space of symmetrized basis functions, thus the reduced probabilities of electric dipole and quadrupole transitions are calculated. More details of the approach are presented in Ref.~\cite{dobrowolski:2016} while the center-of-mass problem has been shown in Ref.~\cite{myAPP2017}.

The interesting aspect of our study is an estimation of the mixing of the K quantum numbers, which stands for a projection of the angular momentum on quantization axis of the intrinsic reference system called K. An interesting outcome of present considerations is a statement that the K quantum number is not conserved within a band and moreover each state contains the mixture of various values of K in uncorrelated percentage.

The paper is organized as follows: Section~\ref{Section.02} gives the details of the collective quadrupole-octupole model and the uniqueness of intrinsic vibrational- rotational Hamiltonian eigensolutions in the laboratory frame.
Section~\ref{Section.03} is devoted to estimations of the negative-parity bands. The rotational properties of this states are also discussed. The branching ratios are compared with experimental data. The article is closed with short summary.

\section{Collective quadrupole-octupole model}
\label{Section.02}
The vibrational-rotational collective bands of the positive or negative parity being the subject of
our interest can be modelled with the use of either even or odd-multipolarity $\alpha_{\lambda\mu}$ 
 deformations, where $\lambda=\{1,2,3,\ldots\}$ and $\mu=-\lambda,-\lambda+1,\ldots, +\lambda$.
In the following applications these deformation parameters become the dynamical collective variables
describing surface vibrations in the intrinsic frame. The variables $\alpha_{\lambda\mu}$ 
are also the spherical components of the irreducible tensor with respect to the $SO(3)$ group, so their properties
are well defined with regard to the group theory formalism. 

The nuclear surface is expanded in the body-fixed reference frame in terms of the orthogonal basis set of 
the spherical harmonics $\{Y_{\lambda\mu}\}$. As shown in \cite{dobrowolski:2016}, the dipole $\alpha_{10}$ and $\alpha_{1\pm 1}$
variables are determined from the condition that the center of mass of the nuclear body is fixed in the beginning of the coordinate system. 

The space spanned by two quadrupole variables, $\alpha_{20}$, $\alpha_{22}=\alpha_{2-2}$
with the conditions $\alpha_{21}=\alpha_{2-1}=0$
defines the body-fixed frame of the discussed model. Therefore, this set together with the full octupole 
$\{\alpha_{3\nu}\}, \nu=0,\pm 1,\pm 2,\pm 3$ complex tensor
and the three Euler angles $\{\Omega\}$ form the twelve-dimensional collective space. However, fixed in this way intrinsic frame is not the
 principal-axes frame of the quadrupole-octupole body but it permits to use the traditional picture of the collective quadrupole motion
 extended by the independent octupole vibrations. 
The calculation of the  matrix elements of the collective Hamiltonian 
and/or any physical observables with a satisfactory accuracy is a serious task in such defined multidimensional space. 

A further limitation of the $\{\alpha_{3\nu}\}$ values to real numbers implies that $\alpha_{3\mu}$ and $\alpha_{3-\mu}$ are mutually dependent.
Obtained in such a way reduction of the collective-space dimensionality to nine dimensions (including Euler angles) allows now for 
an efficient determining the time-consuming multidimensional integrals and consequently, investigate contributions from individual
collective modes.
The use of the so called {\it adiabatic approximation} gives that the vibrational and rotational matrix elements of the Hamiltonian 
can be calculated separately. Moreover, the rotational matrix elements, depending only on the Euler angles, may be calculated analytically.

Actually, the independent vibrational collective variables of the present approach are
 $(\alpha_{20}, \alpha_{22}, \{{\rm Re}(\alpha_{3\nu})\})$ with $\nu$ index running over the positive integers only, i.e. $\nu=0,1,2,3$,
describing the axial, non-axial quadrupole vibrational modes and the four real octupole modes, respectively.
With the above, the nuclear surface can be written as
\begin{eqnarray}
&& R(\vartheta,\varphi)=R_{0}c(\alpha)\bigg[1 + 
    \alpha_{10}Y_{10}(\vartheta,\varphi)+\alpha_{20}Y_{20}(\vartheta,\varphi)+ \nonumber\\
&& 2\alpha_{11}{\rm Re}\big(Y_{11}(\vartheta,\varphi)\big)+2\alpha_{22}{\rm Re}\big(Y_{22}(\vartheta,\varphi)\big)+
   \nonumber\\
&& \alpha_{30}Y_{30}(\vartheta,\varphi)+2\sum_{\mu=1}^{3}\alpha_{3\mu}{\rm Re}\big(Y_{3\mu}(\vartheta,\varphi)\big)\bigg],                                                            
\label{eqn.14}
\end{eqnarray}
where the function $c(\alpha)$ ensures the volume conservation of the deformed body.

The problem of the center of mass shift as a result of the presence of the mass asymmetry in octupole deformed nuclei is widely discussed
in Ref.~\cite{dobrowolski:2016,myAPP2017}.

For a fixed values of the quadrupole deformations $(\alpha_{20},\alpha_{22})$, the conditions ${\rm Im}(\alpha_{3\mu})=0$ on the $\alpha_{3\mu}$ tensor 
cause that a single octupole shape can be obtained with more than one 
 set $(\alpha_{30},\alpha_{31},\alpha_{32},\alpha_{33})$. Generated in this way shapes have, however, different orientations
 with respect to the axes of the laboratory frame.
In order to avoid the non-uniquenesses of the wave functions in the laboratory frame caused
 by this property of the truncated $\alpha_{\lambda\mu}$ collective space along with the definition of the intrinsic frame, 
one should introduce the {\it symmetrization} procedure. 

Briefly, each physical state which 
 describes the system in the laboratory frame should necessarily be invariant with respect to the so-called symmetrization group $\bar{G_s}$.
Such a group is always determined individually, depending on the set of $\alpha_{\lambda\mu}$ variables involved in the model.
For used in this work real octupole variables, the symmetrization group, introduced in Ref.~\cite{dobrowolski:2016}, is 
$\bar{G_s}=\bar{D}_{4y}$ and is lower than the octahedral group in a pure quadrupole Bohr-Hamiltonian model. Its elements (rotations), $\bar{g}$, are: 
($I$, $C_{2x}$, $C_{2y}$, $C_{2z}$, $C_{4y}$, $C_{4y}^{-1}$, $C_{2c}$, $C_{2d}$), where $C_{ni}$ (for $i=\{x,y,z\}$) denote the rotations
about $2\pi/n$ angle around the $i^{th}$-axis.
Finally, for all $\bar{g}\in\bar{G_s}$, the symmetrization condition applied to any collective state $\Psi(\alpha,\Omega)$ reads:
\begin{equation}
\bar{g}\Psi(\alpha,\Omega)=+1\cdot\Psi(\alpha,\Omega).
\label{SymCondStates} 
\end{equation}
The relation (\ref{SymCondStates}) ensures the uniqueness of the Hamiltonian eigensolutions in the laboratory frame. 
In the context of the Hamiltonian-symmetry problem, the symmetrization group $\bar{G_s}$ can be treated as its {\bf minimal}
symmetry group. This implies that both the kinetic and potential components of the full Hamiltonian exploited in this
study have to be, at least, $\bar{G_s}$-invariant.

\subsection{Quadrupole + Octupole Collective Hamiltonian}
\label{Section.02a}

Habitually, a consistent vibrational-rotational collective approach is constructed by defining
the collective Hamiltonian with respect to the laboratory frame spanned by the laboratory
collective variables. In the next step, this Hamiltonian is transformed to the body-fixed frame. 
For the quadrupole collective space a standard kinetic energy term obtained with this prescription results e.g. with the well 
known Bohr Hamiltonian approach.

In contradiction to the above outlined scheme, the used here collective vibrational-rotational Hamiltonian is already written
in the intrinsic frame. Additionally, the above mentioned adiabatic approximation is applied in order to separate
the vibrational and rotational motions. In principle, such separation is possible due to energy scales of both the vibrational 
and rotational modes. It is also assumed that the quadrupole and octupole vibrational modes
are totally decoupled in the kinetic-energy term. This accelerates the numerical calculations by a factor equal to the number of
mesh points of the quadrupole space $\{\alpha_{20},\alpha_{22}\}$, i.e. about $2\times 10^3$.

Therefore we calculate two independent mass tensors: first for pure quadrupole motion,
with the octupole deformation corresponding to the potential-energy minimum, and the second, referring to the octupole motion only, 
for which the quadrupole deformations describe the ground state shape. 

This simplification leads to a quantized realistic quadrupole-octupole-vibrational Hamiltonian with deformation-dependent
 inertia parameters 
\begin{eqnarray}
 && {\mathcal H}_{coll}(\alpha_2,\alpha_3,\Omega)=\frac{-\hbar^2}{2}\bigg\{\nonumber\\
 && \frac{1}{\sqrt{\vert B_2\vert}}\sum\limits_{\nu\nu^{\prime}=0}^2
  \frac{\partial}{\partial \alpha_{2\nu}}
   \sqrt{\vert B_2\vert} \big[B_2^{-1}\big]^{\nu\nu^{\prime}}
    \!\!\!\frac{\partial\;\;}{\partial \alpha_{2\nu^{\prime}}}+\nonumber\\
 &&  \frac{1}{\sqrt{|B_3|}}\sum\limits_{\mu\mu^{\prime}=0}^3
   \frac{\partial}{\partial \alpha_{3\mu}}
   \sqrt{|B_3|} \big[B_3^{-1}\big]^{\mu\mu^{\prime}}
   \!\!\!\frac{\partial\;\;}{\partial \alpha_{3\mu^{\prime}}}\bigg\}+ \nonumber\\
&& \hat H_{rot}(\Omega) + \hat V(\alpha_2,\alpha_3),
\label{Hcoll}                                                  
\end{eqnarray}
where $\alpha_2$ and $\alpha_3$ describe subspaces of the quadrupole and octupole variables with metrics 
$B_2(\alpha_2)$, $B_3(\alpha_3)$ given in this approach as the quadrupole and octupole microscopic mass tensors, respectively.
Quantities $|B_2|={\mathrm{det}}(B_2(\alpha_2))$ and $|B_3|={\mathrm{det}}(B_3(\alpha_3))$ stand for the square roots of the metric-tensor determinants. 
These microscopic mass tensors are determined using {\it cranking} method of Ref.~\cite{Inglis}. Its covariant component, $B_{\lambda\nu,\lambda\nu'}$, 
for $\lambda=2$ or $\lambda=3$ and indices $\nu>0$ is given by the expression
\begin{eqnarray}
B_{\lambda\nu,\lambda\nu'}(\{\alpha_{\lambda\mu}\})&=&\sum\limits_{kl}
\frac{\langle \phi_k|\frac{\partial\hat H_{sp}}{\partial\alpha_{\lambda\nu}} |\phi_l\rangle\, 
      \langle \phi_l|\frac{\partial\hat H_{sp}}{\partial\alpha_{\lambda\nu'}}|\phi_k\rangle}
        {(E_k+E_l)^3} \times\nonumber\\
      &&\big(u_k\,v_l+v_k\,u_l\big)^2,
\label{cran_mass}\end{eqnarray}
where the double sum runs over the full set of the BCS quasi-particle (including time-reversed) states, obtained out of the eigensolutions of used mean-field
Hamiltonian $\hat H_{sp}$ and chosen pairing model. Quantities $v_n$ are the occupation probability amplitudes of the $n^{th}$ quasi-particle
state while $u_n$ is given by the normalization relation $u_n^2=1-v_n^2$.  In the denominator of Eq.~(\ref{cran_mass}), $E_k$ and $E_l$ are the quasi-particle
energies of $k^{th}$ and $l^{th}$ states.

In this work, contrarily to applied nowadays self-consistent methods, an effective approximation to generate the collective potential in the six-dimensional space of $\{\alpha_2,\alpha_3\}$ variables
is still widely applied macroscopic-microscopic model. This model, for a reasonable choice of the mean-field potential, pairing interaction and the smooth 
liquid-drop energy contribution, is able to produce reliable estimates of potential energy surfaces $\hat V(\alpha_{2\nu},\alpha_{3\mu})$. 
Within this studies we use the Woods-Saxon potential \cite{wosax} with the so called {\it universal}
set of parameters \cite{cwiok87} (refitted to the newer single particle data of \cite{nicodok}) which delivers the single-particle energies and eigenstates for a given mean-field deformation.
Both these quantities are the initial quantities to the calculations of the quantum shell and pairing energies as well as mass parameters via Eq.~(\ref{cran_mass}).

The shell-energy correction arising due to the shell structure of a nuclear system is calculated using the standard Strutinsky
approach of $6^{th}$ order \cite{strut66,strut67,strut68}. Furthermore, for the pairing energy the particle number projected BCS approach \cite{BCSbs63,bolstr} is used.
Eventually, the leading liquid-drop energy term is developed here by the Lublin-Strasbourg Drop formula (LSD) \cite{KPD03} which permits to
successfully reproduce fission barriers of actinides, see e.g. \cite{JDu04}.

\subsection{Rotational Hamiltonian}
\label{Section.02b}

Due to significantly different energy regimes of the vibrational and rotational modes, they are here totally decoupled. Hence, the rotational term $\hat{H_{rot}}(\Omega)$ depends only on the Euler angles, and parametrically, the 
static nuclear deformation, now corresponding to the equilibrium point. Since, as mentioned in Section \ref{Section.02}, the
rotational Hamiltonian has to be invariant with respect to the symmetrization group $\bar{G_s}$, we construct it using the irreducible (spherical) 
tensors of $\bar{G_s}$ group, $\hat {\mathcal T}_{\lambda\mu}(n;\lambda_2=2,\lambda_3=3,...,\lambda_{n-1}=(n-1))$, as done e.g. in Refs.~\cite{gen_Hrot,gen_Hrot2,dobrowolski:2016}. 

The rotor Hamiltonian $\hat{\mathcal H}_{rot}$ of given symmetry and multipolarity $\lambda$ can be built as the linear combination of $\hat {\mathcal T}$ over indices
$\lambda$ and $\mu$ with $n=\lambda$ and the term $T_{00}(n=2)$ as 
\begin{eqnarray}
\hat{\mathcal H}_{rot}= \sum\limits_{\lambda=0}^{\lambda_{max}} \sum\limits_{\mu=-\lambda}^{\lambda}\,
c_{\lambda\mu}\,\hat{\mathcal T}_{\lambda\mu} + c_{00}T_{00}(n=2).
\label{genHrot}\end{eqnarray}
The upper limit of multipolarities $\lambda_{max}$ is, in general, arbitrary. In this work we limit ourselves to $\lambda_{max}=2$

The coupling constants 
$c_{00}$, $c_{20}$ and $c_{22}$ are functions of the moments of inertia as
\begin{eqnarray}
&&c_{00}=-\frac{1}{\sqrt{12}}\left(\frac{1}{I_x} + \frac{1}{ I_y} + \frac{1}{ I_z}\right), \nonumber\\             
&&c_{20}= \frac{1}{\sqrt{6 }}\left(\frac{1}{I_z} - \frac{1}{2I_x} - \frac{1}{2I_y}\right), \nonumber\\
&&c_{22}= \frac{1}{4}\left(\frac{1}{I_x} - \frac{1}{I_y}\right),
\label{coupl-const}\end{eqnarray}
where $I_x$, $I_y$, $I_z$ are the microscopic nuclear moments of inertia with regard to $Ox$, $Oy$ and $Oz$ axes, respectively, obtained in the cranking approximation.

If the $\bar{D}_{4y}$-symmetric rotor Hamiltonian $\hat H_{rot}(\Omega)$ of Eq.~(\ref{Hcoll}) is needed, the quadrupole coupling constants entering Eq.~(\ref{genHrot}) 
are related by $c_{22}\approx c_{20}/0.8165$.


Following the symetrization idea, the
basis in which the full collective Hamiltonian (\ref{Hcoll}) is diagonalized, contains functions symmetrized with respect 
to the intrinsic group $\bar{D}_{4y}$. Remind that the intrinsic group, by definition, acts in the intrinsic collective space containing Euler angles.

In numerical calculations it is very convenient to use the projection operator formalism which defines the projection of an initial wave function onto the selected irreducible
representation of the symmetry group. If one chooses in particular the scalar (A1) representation of the symetrization $\bar{G}_s$ group, such a procedure is equivalent to the symmetrization condition (\ref{SymCondStates}).

Applying the explicit form of the projection operator on the six-dimensional ''shifted'' harmonic oscillator solution combined with the appropriate Wigner function, one gets the symmetrized basis function 
\begin{widetext}
\begin{eqnarray}
\Psi_{k;JM\kappa}^{(\pm)}&=&\hat P^{(A1)}\,\Psi_{k;JMK\pi}=\sqrt{2J+1}\sum\limits_{K=-J}^J D^J_{\kappa K}(g) \frac{1}{8}
\sum\limits_{i=1}^{8} \,
                 u_{n_{20}}(\eta_{20},\hat{\bar{g_i}}\alpha_{20}-\mathring{\alpha}_{20}) u_{n_{22}}(\sqrt{2}\eta_{22},\hat{\bar{g_i}} \alpha_{22}-\mathring{\alpha}_{22})\nonumber\\
              &&
                 u_{n_{30}}(\eta_{30},\pm\hat{\bar{g_i}} \alpha_{30}-\mathring{\alpha}_{30})
	         u_{n_{31}}(\sqrt{2}\eta_{31},\pm\hat{\bar{g_i}} \alpha_{31}-\mathring{\alpha}_{31})                                                                
                 u_{n_{32}}(\sqrt{2}\eta_{32},\pm\hat{\bar{g_i}} \alpha_{32}-\mathring{\alpha}_{32})\nonumber\\
               &&  u_{n_{33}}(\sqrt{2}\eta_{33},\pm\hat{\bar{g_i}} \alpha_{33}-\mathring{\alpha}_{33}),                 
\label{proj_Psi}\end{eqnarray} 
\end{widetext}
where the set of all elements $\hat{\bar{g}}_i$ forms the symmetrization group. 
The parameters $\mathring{\alpha}_{2\nu}$ and $\mathring{\alpha}_{3\mu}$ describe the position of the potential-energy well minimum. Studying the potential-energy maps
of Section (\ref{Section.02c}) we can conclude in advance that $\mathring{\alpha}_{20}=0.25$, $\mathring{\alpha}_{22}=0$ and all $\mathring{\alpha}_{3\mu}=0$.

The functions of the positive parity (+) or negative (-) parity are obtained as the linear combinations of those of Eq.~(\ref{proj_Psi}) as
$\frac{1}{2}[\Psi_{k;JM\kappa}^{(+)}+\Psi_{k;JM\kappa}^{(-)}]$ and $\frac{1}{2}[\Psi_{k;JM\kappa}^{(+)}-\Psi_{k;JM\kappa}^{(-)}]$, respectively. 
%
%

\subsection{Collective Potential}
\label{Section.02c}


\begin{figure}[ht!]
   \begin{center}
      \includegraphics[scale=0.3]{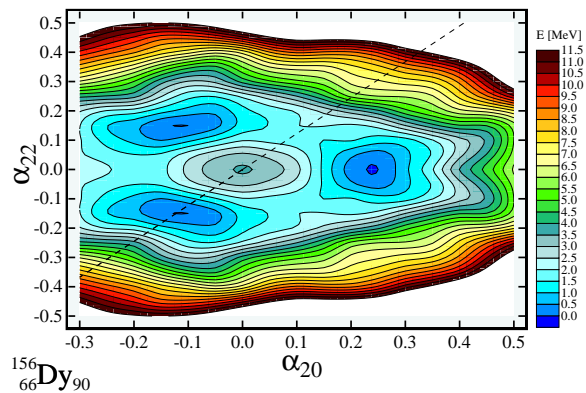}
     \caption{The potential energy of the $^{156}$Dy in the quadrupole plane 
      ($\alpha_{20},\alpha_{22}$). } \label{fig.05}                                                             
   \end{center}
\end{figure}

As already mentioned, the estimates of the total potential energy of the deformed nucleus are done within the phenomenological mean-field approach,
 known as the macroscopic-microscopic method of Strutinsky. 
In this method, as commonly known, the macroscopic energy term given usually by the liquid-drop type formula is modified 
by the microscopic, shell and pairing energy corrections, describing quantum effects in a nucleus. However this kind of approach has been applied for more
 than five decades now, it is still a powerful and successful method, well suited particularly to large scale calculations, able to produce results close to the 
experimental data. The details of this kind of calculations and corresponding results are presented e.g. in \cite{dudek2004,dudma2004,dudma2005,mazdu2005,mazdj2005}.

The geometry of the potential energy surfaces in the vicinity of the equilibrium state of cold, medium mass nuclei generated, for example,  
by the Lublin-Strasbourg Model (LSD) \cite{KPD03} is very similar to this, obtained from other competitive macroscopic models. 
\begin{figure*}[ht!]\hskip -0.5cm
      \includegraphics[scale=0.29]{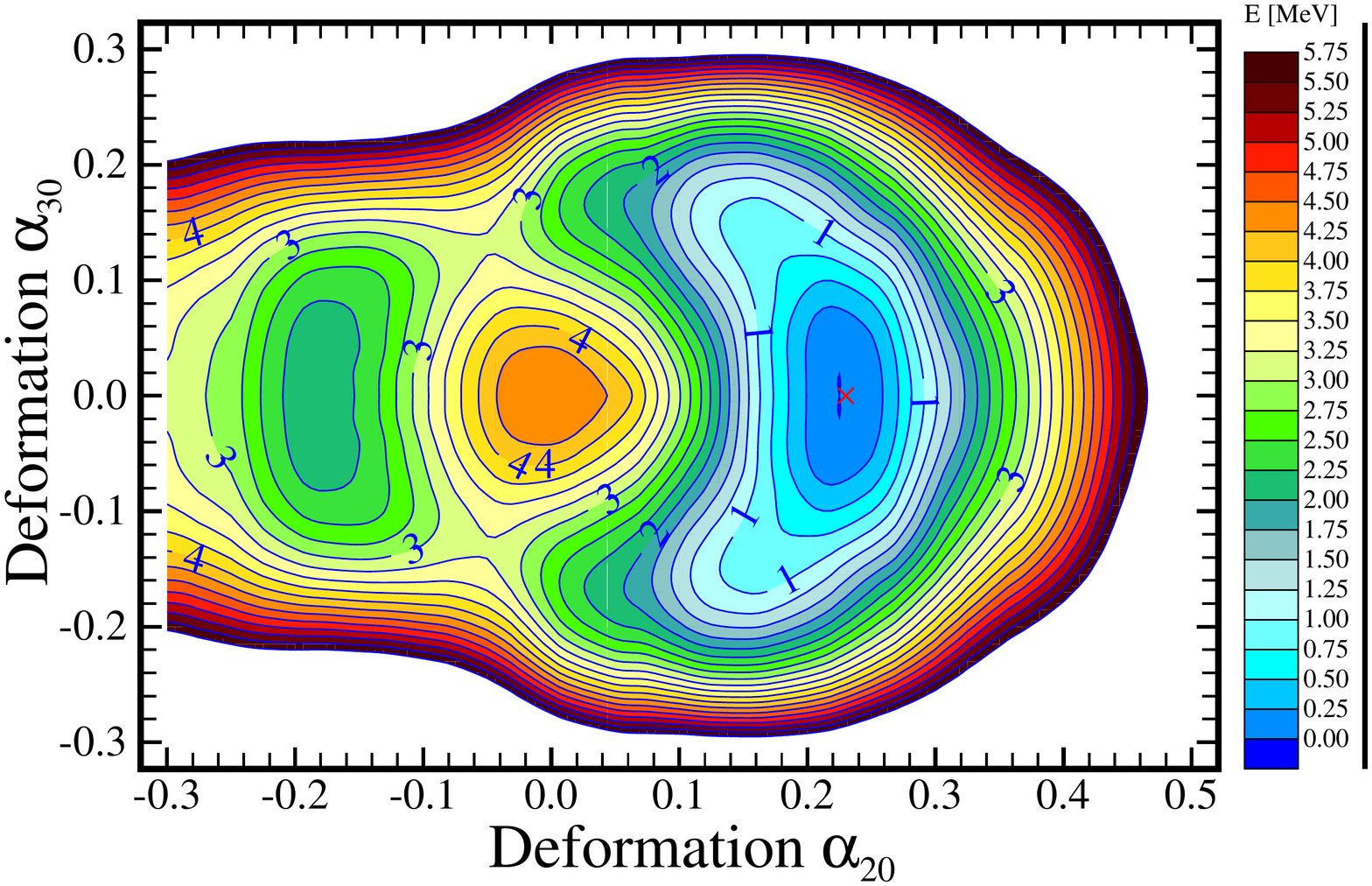}\hskip -0.5cm
      \includegraphics[scale=0.29]{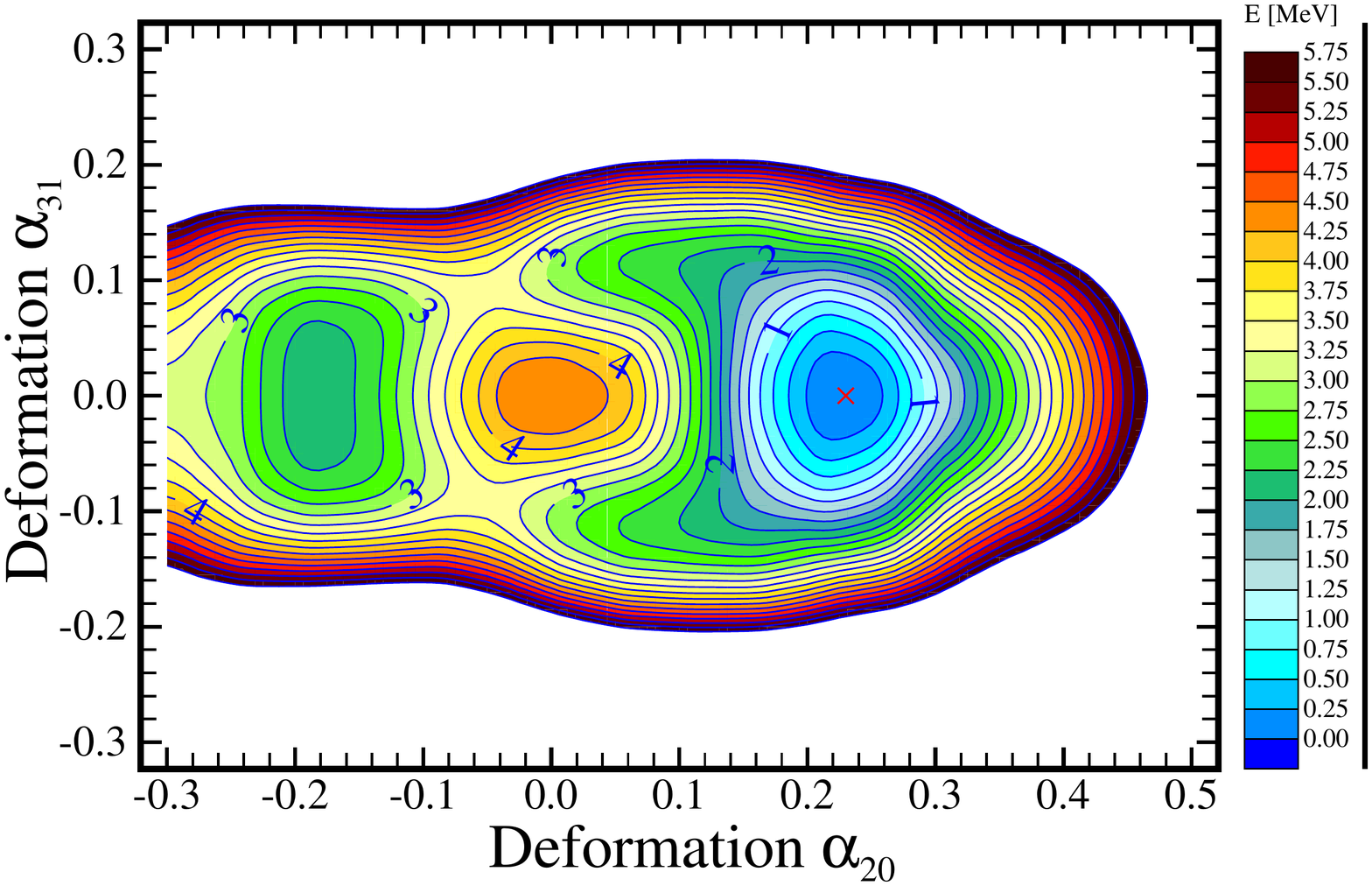}\\ \vskip -1.5cm\hskip -0.5cm
      \includegraphics[scale=0.29]{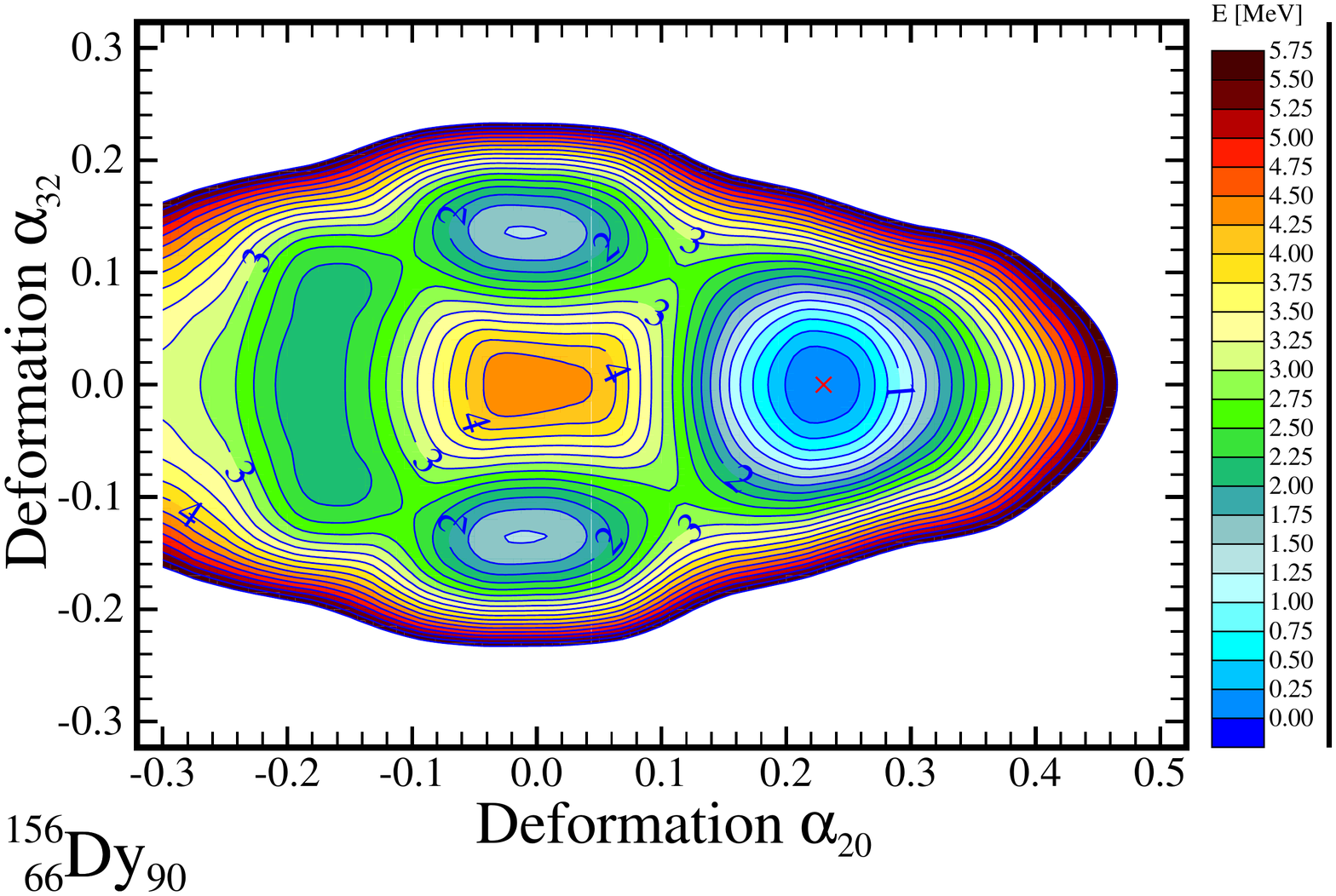}\hskip -0.5cm
      \includegraphics[scale=0.29]{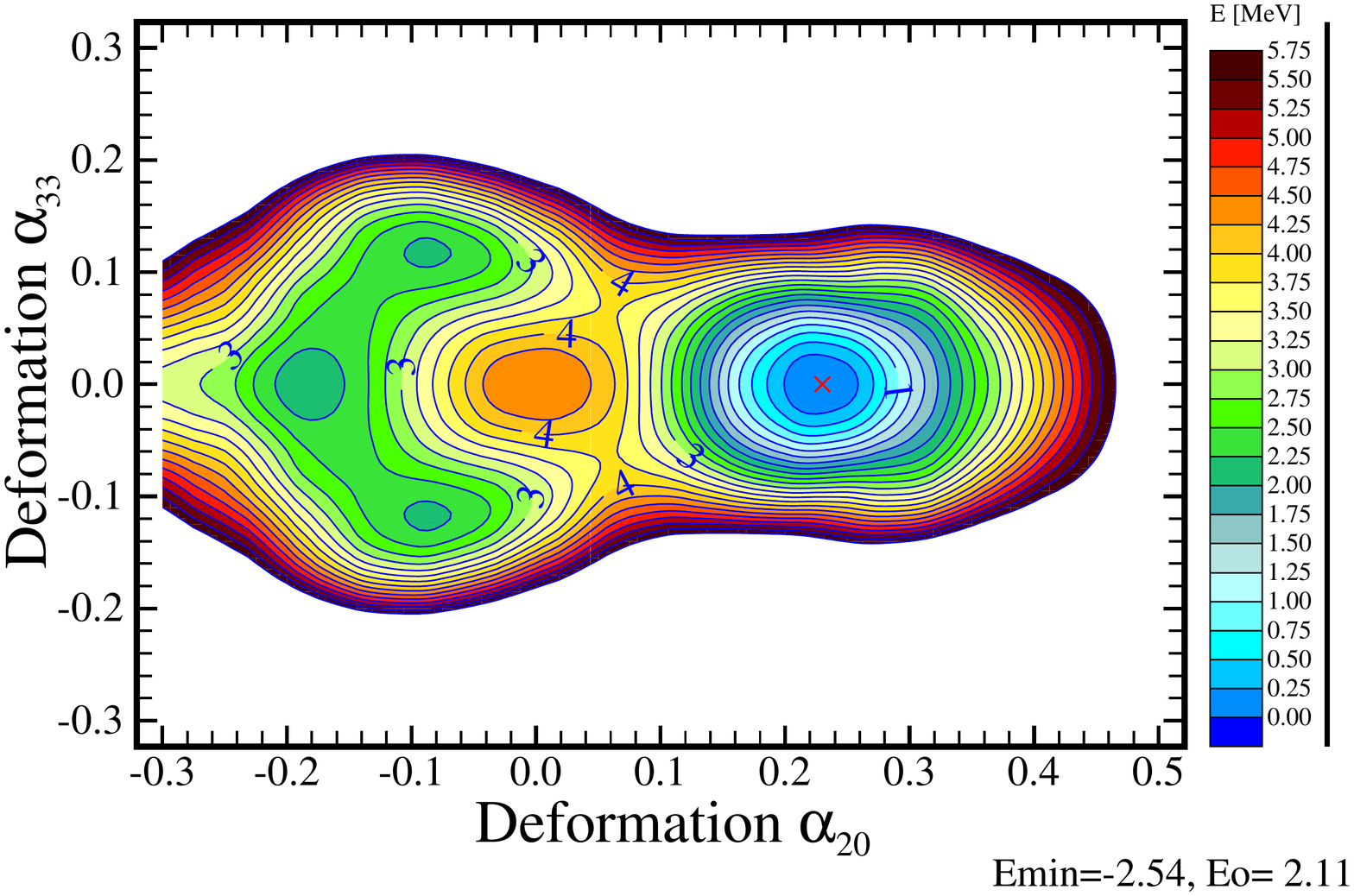}
    \caption{Potential energy maps generated for the quadrupole $(\alpha_{20},\alpha_{22}=0)$ versus octupole ($\alpha_{30}$, $\alpha_{31}$, $\alpha_{32}$, $\alpha_{33}$)
             deformations. }
\label{fig.06}
\end{figure*}

The microscopic energy correction is defined as the sum of the shell and pairing energy corrections to the smoothly changing liquid drop energy. 
The shell energy is obtained from the Strutinsky method developed in Refs.~\cite{strut66,strut67,strut68}. For the pairing energy \cite{BCSbs63,bolstr}
as the difference between the sum of the single particle energies and the energy of the pair correlations \cite{krappe:1979},
the particle-number projected (PNP) pairing model \cite{bolstr} within the standard BCS framework is applied.

The numerical calculations of the total
collective potential entering Eq.~(\ref{Hcoll}) are performed in the six-dimensional mesh of vibrational collective variables: $\{\alpha_{20}, \alpha_{22},
\alpha_{3\nu}, \nu=0,1,2,3\}$ for $^{156}$Dy nucleus.

The ranges of nuclear deformation parameters as well as the corresponding mesh steps 
$\Delta\alpha_{\lambda\mu}$ are listed below: 
\begin{eqnarray}
 \alpha_{2\nu}\epsilon (-1.0;1.0), \quad&\Delta \alpha_{2\nu}= 0.05, \quad &\nu=0,2  \nonumber\\
 \alpha_{3\mu}\epsilon (-0.3;0.3), \quad&\Delta \alpha_{3\mu}= 0.1, \quad &\mu=0,1,2,3   \end{eqnarray}
which gives the mesh of about two millions points, describing various quadrupole-octupole nuclear shapes.

Fig.~\ref{fig.05} displays the total energy map as function of the quadrupole 
($\alpha_{20},\alpha_{22}$), putting the other four deformation parameters to zero. 
The equilibrium energy minimum corresponding to the quadrupole axial (prolate) shape of $^{156}$Dy is visible. 
The straight dashed line of Fig.~\ref{fig.05} on ($\alpha_{20},\alpha_{22}$)  
cross-section separates the quadrupole configurations which are identical with respect to the $\bar{D}_{4y}$ symmetrization group. We observe the ground state energy well occuring in the three ($\alpha_{20},\alpha_{22}$) quadrupole configurations.

The problem of the ''repeatability'' of the nuclear shapes as a results of the symmetrization with respect to the octahedral and $\bar{D}_{4y}$ groups, is widely discussed in 
Ref.~\cite{dobrowolski:2016} and references therein. 
Now, we want to recall that, in particular, the resulting Strutinsky potential energy as a function of the quadrupole and octupole deformation 
is invariant with regard to the symmetrization group $\bar{G}_s$. This property is true since the macroscopic liquid drop contribution as well as the shell and pairing microscopic
energy corrections depend only on the shape of the nuclear surface defined by (\ref{eqn.14}).
This means that for a fixed quadrupole deformation, a single octupole shape for all $\alpha_{3\mu}^{(0)}\neq 0$, can be obtained by using eight different deformation-parameter combinations. 
 In general, the identical quadrupole-octupole shape for the $\bar{D}_{4y}$ symmetrization group are expected to show up, at maximum, $2\times 8=16$ sixteen times in the full ($\alpha_2,\alpha_3$) space. 
Otherwise, if it happens that all $\alpha_{3\mu}^{(0)}=0$, such a shape appears, in fact, three times. Please remind that in this particular case, the true symmetrization group is the octahedral, not $\bar{D}_{4y}$, group.

The dependence of the total potential energy on the quadrupole $\alpha_{20}$ and octupole $\alpha_{3\mu}$ degrees of freedom is shown in Fig.~\ref{fig.06}.
Projections of full PES into axial quadrupole and selected octupole deformation parameters space permit to trace the features of the global and local energy minima, such as 
their positions and depths. 

\begin{figure*}[hbt!]\hskip -0.5cm
   \begin{center}
      \includegraphics[scale=1.0]{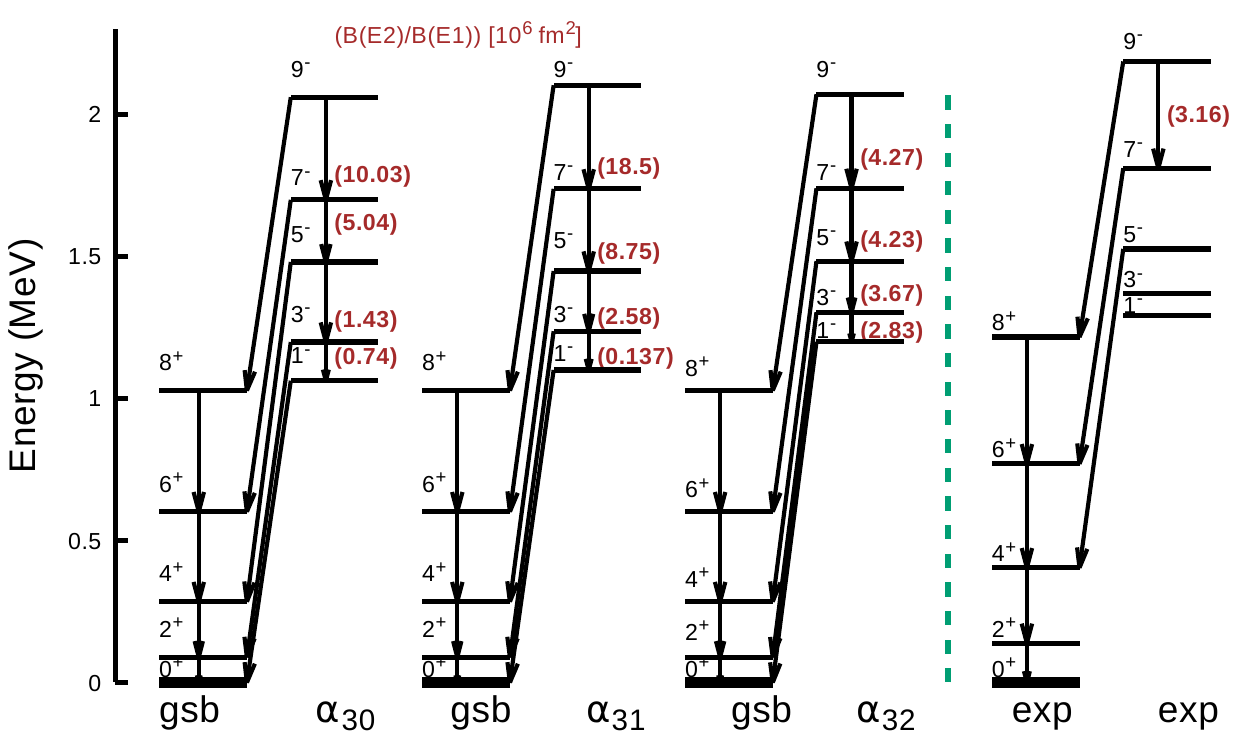}
     \caption{The predicted ground state and negative-parity bands of the $^{156}$Dy. Arrows mark the E1 and E2 transitions. The branching ratios B(E2)/B(E1) are written in parenthesis. } \label{fig.08}                                                             
   \end{center}
\end{figure*}

The total energy maps projected on ($\alpha_{20}$, $\alpha_{3\mu}$) plane show subtly pronounced two identical minima for $\alpha_{20}<0$ and $\alpha_{22}=0$.
Figure~\ref{fig.06} displays that the ground-state well, which appears for octupoles $\alpha_{3\nu}=0$, is the only stable configuration for $^{156}$Dy nucleus.
Thus, the above written arguments lead to the conclusion that one should obtain exactly two additional ''copies'' of this minimum, both again for 
$\alpha_{3\nu}= 0$. In Fig.~\ref{fig.05}, this minima are visible for $\alpha_{20}$ slightly lower than zero and $\alpha_{22}=0$.

\section{Results}
\label{Section.03}

The here discussed model offers the positive and negative-parity collective vibrational-rotational states based on $\alpha_{2\mu}$ and $\alpha_{3\nu}$ one-phonon excitations.
\subsection{Negative parity bands}
\label{Section.03a}

 The negative-parity states are created in the potential energy well based on the quadrupole-deformed ground-state configuration with quadrupole deformation
$\alpha_{20}=0.25$ and $\alpha_{22}=0.0$. By consequence, the resulting octupole negative-parity states have significant static quadrupole deformation producing
large $B(E2)$ intra-band transition probabilities. As deduced from potential energy plots of Fig.~\ref{fig.06}, the octupole vibrations (with multipolarity $\lambda=3$)
are performed around pure quadrupole shapes, i.e. $\alpha_{3\nu}=0$, $\nu=0,1,2,3$.

As seen in Fig.\ref{fig.08}, the negative-parity bands, having as the band-head axial and non-axial $\alpha_{30}$ and $\alpha_{31}$ one-phonon vibrational excitations,
 are shifted each other in energy hardly by about 70 keV
whereas the band built on the tetrahedral $\alpha_{32}$ phonon lies higher by approximately 150 keV. The band built at the $\alpha_{33}$ one-phonon state is too
high in energy compared to the previous bands, so are not considered in this study.

Due to energetically proximity of the mentioned three of all four octupole bands, the photon energies of the dipole inter-band transitions (Tab.~\ref{tab.01}), $E_{\gamma}(\lambda=1)$, 
vary within an interval $\pm 0.1$~MeV which is even less than the order of typical discrepancy between the experimental results and theoretical predictions in
 up-to-date models.

\begin{table}[hbt!]
\caption{The $\gamma$-rays energies predicted theoretically ($E_{\gamma}^{th}$) and measured experimentally 
($E_{\gamma}^{exp}$) at Gammasphere on $^{156}$Dy \cite{Har17}. The energies of the 
band-heads of the 
negative-parity states characterized by various types of the one-phonon octupole excitations are shown.}\label{tab.01}\vspace{0.2cm}
\begin{tabular}{cccccc}
\hline
Transition & $E_{\gamma}^{\alpha_{30}}$&$E_{\gamma}^{\alpha_{31}}$&$E_{\gamma}^{\alpha_{32}}$&$E_{\gamma}^{\alpha_{33}}$&$E_{\gamma}^{exp}$\\
           &  keV&keV&keV&keV&keV\\
\hline
$(3^-\to 1^-)$& 136&136&104&104& \\
$(5^-\to 3^-)$& 212&212&180&180& \\
$(7^-\to 5^-)$& 289&289&247&152&  \\
$(9^-\to 7^-)$& 361&365&331&327& 376 \\
\hline
$(5^-\to 4^+)$& 1115 & 1153&1188&1621&1121 \\
$(5^-\to 6^+)$& 809  & 847 &882 &1315&755 \\
$(7^-\to 6^+)$& 1098 & 1136&1138&1577&1039\\
$(7^-\to 8^+)$& 671  & 709 &711 &1150&594\\
$(9^-\to 8^+)$& 1032 & 1074&1042&1477&971\\
\end{tabular}
\end{table}

One is therefore not able to surely indicate at this stage which of these negative-parity bands is the best candidate to reproduce the experimental collective 
band refereed to as ''band 2'' in \cite{dobrowolski:2016}.
The intra band $B(E2)$ values in the band built on axial $\alpha_{30}$ mode are in general, lower by approximately 40\% than in the ''tetrahedral'' band what is somehow
in contradiction to early simplistic approaches to identify the tetrahedral symmetry.
As commonly known, their values are predominantly determined by the quadrupole moment of the band-head and the rotational structures in terms of the K-number distribution of the initial and final collective states. As explained in \cite{dobrowolski:2016}.

\subsection{Rotational properties of states}
\label{Section.03b}

\begin{figure}[ht!]
   \begin{center}
      \includegraphics[scale=0.7]{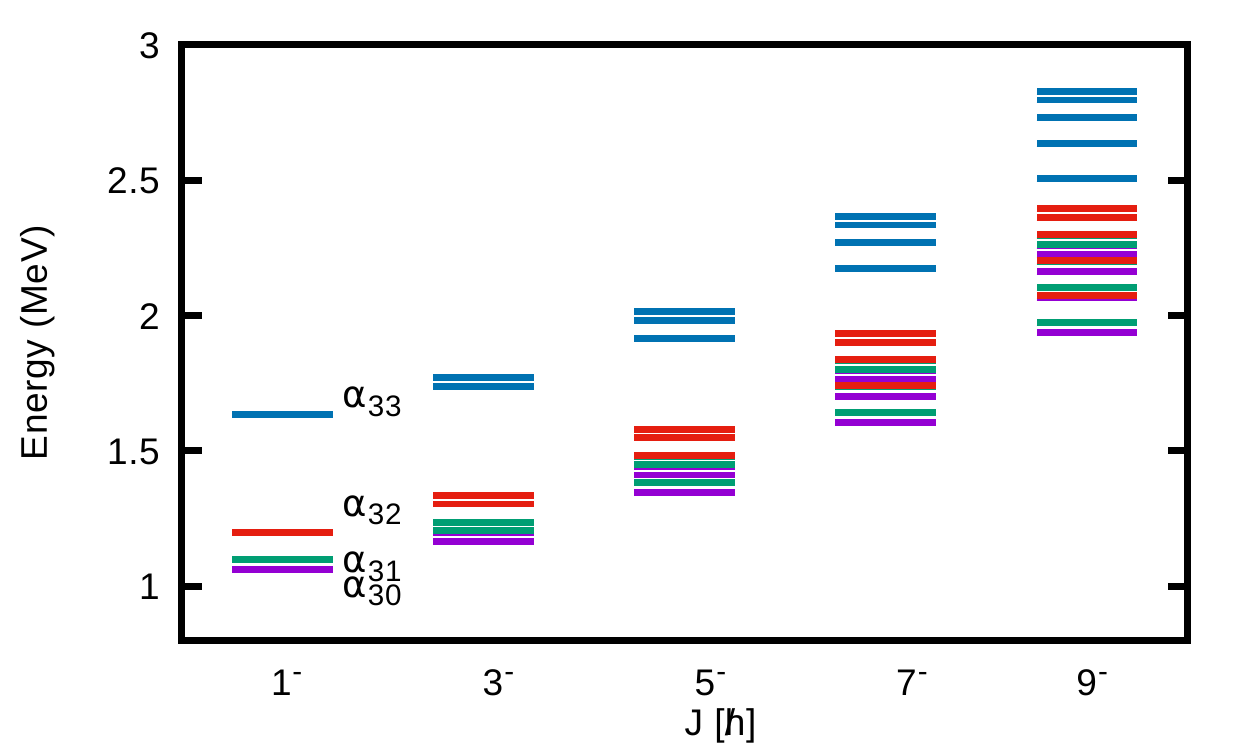}
     \caption{(Color on-line) The negative-parity states build on octupole axial and non-axial one-phonon excitations of the $^{156}$Dy. Colors mark (J+1)/2 states with different $\kappa$ numbers for given spin J.} \label{fig.07}                                                             
   \end{center}
\end{figure}

Each vibrational-rotational negative-parity state of given spin $J$ (in this work, $0\leq J \leq 9$), characterized by a given type of octupole
 excitation and the number of excited phonons (1 or 3), can occur in $2J+1$ configurations as shown in Fig.~\ref{fig.07}. They are described by
specific combinations of rotational basis functions given as complex conjugated Wigner functions $D_{MK}^{J\star}(\Omega)$ with given $K$ numbers. 
These combinations labeled here by $\kappa$ number ($-J\leq \kappa \leq J$) are fixed
 to ensure that the collective states are symmetrized with respect to the $D_{4y}$ group, widely discussed in \cite{dobrowolski:2016},
which conserves the body-fixed frame the additional conditions on the octupole variables ${\rm Im}(\alpha_{3\nu})=0$ during the collective motion.
In brief, the symmetrization condition requires that each collective state as the eigensolution of the collective Hamiltonian of Eq.\ref{Hcoll} 
has to be invariant with respect to the symmetrization group. 
Obtained in this way solutions are unique in the laboratory frame and deserve
to be called as {\it physical} states. 
Let us also observe that the rotational Hamiltonian (\ref{genHrot}), by construction, 
contains, besides the constant term $T_{00}$, the angular momentum operators in power two only, thus it is time-reversal invariant. 
By the fact that the vibrational Hamiltonian term defined in Section \ref{Section.02a} is obviously
 invariant with regard to the time-reversal operation, one concludes that the full Hamiltonian keeps this symmetry.

In general, as written in Ref.~\cite{dobrowolski:2016} the states with negative $\kappa$ value are lineary dependent on those with positive $\kappa$.
This means that they contains the identical, real combinations of $K$'s (differing only with the sign of $K$) and therefore can be considered as mutualy
 time-reversed states. Consequently, it is sufficient to solve the Hamiltonian eigenproblem within the subset of basis functions with e.g. $\kappa > 0$.
Finally, each resulting eigenstate has to be treatad as doubly degenerated. 

\begin{figure}[ht!]
   \hskip -0.5cm
      \includegraphics[scale=0.36]{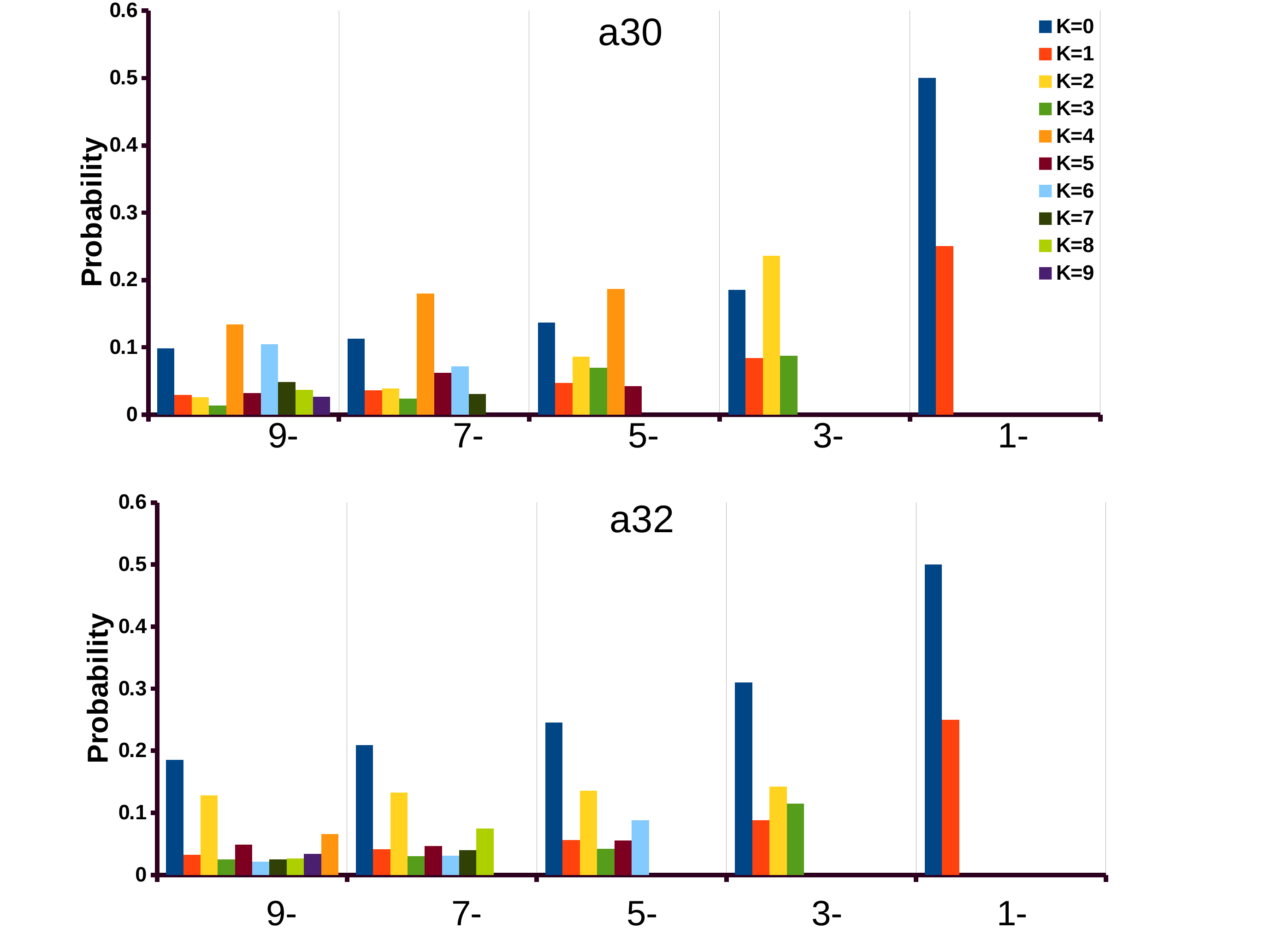}
     \caption{The total energy of the $^{156}$Dy in the quadrupole plane 
      ($\alpha_{20},\alpha_{22}$). } \label{fig.09}                                                             
\end{figure}

Figure \ref{fig.09} presents a distribution of rotational basis states of given $K$ in the collective eigensolutions of spin $J$ belonging to the above
 discussed three model bands based on $\alpha_{30}$, $\alpha_{31}$ and $\alpha_{32}$ one-phonon excitations.
One observes that in the first, axial-octupole, band the dominating rotational components are characterized by $K=0$ and $K=4$ while in the second, 
non-axial band, preferable components are with $K=0$ and $K=2$.
One therefore deduces that none of the considered theoretical bands characterized by one-phonon $\alpha_{3\nu}$ excitation has well fixed
spin projection K number in the chosen intrinsic frame. Let us remind that the conditions imposed on quadrupoles, i.e.
 $\alpha_{21}=\alpha_{2-1}=0$ and $\alpha_{22}=\alpha_{2-2}$ to define the body-fixed coordinate system are identical as in the Bohr-Hamiltonian model
of e.g. Ref.\cite{Pro09}.

Seemingly, the mixing of the $K$ numbers within a given octupole band is caused by relatively high-order symmetrization group $\bar{D}_{4y}$
 which combines rotational basis states of different $K$, as defined in Eq.~\ref{proj_Psi}. 
As demonstrated in Ref.\cite{kaz-grupy-wewnetrzne}, the use the most general complex space of all octupole variables $\alpha_{3\mu}\in {\mathcal C}$, instead of real ones
 exploited in this study,
 leads to the octahedral symmetrization group. As known from the studies of quadrupole bands within in the Bohr Hamiltonian approach, this group similarly
 as the $\bar{D}_{4y}$ one, mixes the rotational states of different $K$'s as well. 
As an exception, one can imagine an unrealistic collective model with the symmetrization group composed only of rotations about the intrinsic $OZ$ axis, 
which would keep the K number unchanged within the whole collective band. In addition, any
group composed of $C_{nz}$ rotations about $2\pi/n$ angle, where $n\in {\mathcal N}$ and being a subgroup of the $SO(2)$ impose the condition for variables $\alpha_{\lambda\rho}$, 
that $\rho/n$ should be an integer number. 
This implies that for $n \le 3$ some deformations $\alpha_{2\mu}$ and $\alpha_{3\nu}$ are eliminated from the model or, in case $n \ge 3$, only
axial deformations $\alpha_{\lambda 0}$ are left. Another non-physical symmetrization group could be $C_{2y}$ group which contains the only rotation about $\pi$ angle 
with respect to the $OY$ axis. Remind that this particular rotation does not mix the rotational contributions with different K.

On the other hand, a realistic three conditions imposed on the quadrupole and/or octupole variables
 which allow to fix the intrinsic frame in the unique way always lead to symmetrization groups possessing more than one rotation axis. 
One therefore deduce that mixing of $K$ numbers within a collective band is not at all the effect of the presence of limiting conditions imposed
on the collective variables or computational artefacts, hence should be rather treated as true physical property of collective bands.
\subsection{Strenght of $B(E\lambda)$ transitions}
\label{Section.03c}

According to the experimental indications, the ground state well in this 
nucleus is strongly quadrupole deformed. This means that in the equilibrium state octupole degrees of freedom are, 
in the first approximation, not excited. It implies that in the function (\ref{proj_Psi})
$\sum\limits_{\rho=0}^3\,n_{3\rho}=0$ whereas $n_{20}$ and $n_{22}$ values are assumed to be $0,1,2,3$.
For the negative-parity states, on the contrary, $n_{20}=0$ and $n_{22}=0$ while in the octupole part of this function, 
$\sum\limits_{\rho=0}^3 n_{3\rho}=1 {\mathrm or} 3$. Due to the parity property, even-phonon numbers in the right hand side of the previous condition
are not allowed. 

An additional problem is to fix the values of ''shift'' parameters $\mathring{\alpha}_{2\mu}$ and $\mathring{\alpha}_{3\nu}$ in Eq.~(\ref{proj_Psi}) for 
$\lambda=2,3$ and $-\lambda\leq \mu \leq\lambda$. These parameters cause that all the basis functions (\ref{proj_Psi}) are
centered over the potential energy well which minimum in our six-dimensional deformation space is in the point
$(\alpha_{20},\alpha_{22},\alpha_{30},\alpha_{31},\alpha_{32},\alpha_{33})=(0.25,0.0,0.0,0.0,0.0,0.0)$. 

The last group of basis parameters in function (\ref{proj_Psi}) are the so-called ''width'' parameters, $\eta_{\lambda\mu}$. 
Certainly, for the incomplete basis set, the parameters $\eta_{\lambda\mu}$ introduced in Eq.~(\ref{criter_eta}), are absolutely crucial. 
We have decided that the optimal values of these parameters should correspond to the minimal energy of the
ground state, i.e.
\begin{equation}
E_{gs}=E_{gs}(\{\eta_{\lambda\mu}\})={\rm min}.
\label{criter_eta}\end{equation}
In general, the reduced transition probability of the electric $E\lambda$ transition finally reads
\begin{eqnarray}
B(E\lambda, J\to J')= \frac{\left|\Bra{\Phi_{J'\pi'\kappa'}}
|Q^{(lab)}_\lambda |\Ket{\Phi_{J\pi \kappa}}\right|^2}{2J+1}.
\end{eqnarray}
where $J$, $\pi$ and $\kappa$ are quantum numbers of the model, the $Q^{(lab)}_\lambda$ is the transition operator defined in laboratory frame as in the \cite{JEi87}.

Since, as above mentioned, in the case of $\bar{D}_{4y}$ symmetrization group, $K$ is not conserved within the band, one may in fact, construct more than one band of collective states
described by the same vibrational structure and different combinations of $K'$s. 
In order to assemble a band of theoretical eigensolutions one has admitted that within this sequence the 1-phonon vibrational structure is conserved and, in parallel, the intra-band $B(E2)$ values have to lower monotonically with lowering spin. The strengths of the quadrupole $B(E2)$ matrix elements are determined both by the intrinsic vibrational (quadrupole moments) and rotational (appropriate Clebsch-Gordan coefficients) properties of the state in the same footing.
 
Presented here arguments lead, in principle, to rather qualitative conclusions. The authors do not claim that used model perfectly reproduces the absolute values of energies $E_{\gamma}$ (Tab.~\ref{tab.01}) and transition probabilities $B(E1)$ and $B(E2)$ (Tab.~\ref{tab.01}).

\begin{table}[h!]
\caption{The predicted $E2$ transition probabilities for
negative-parity bands. The $E1$ probabilities for transition between negative parity and ground-state bands are given for $^{156}$Dy. The mode of octupole 
         excitation is shown.}\label{tab.02}
\begin{tabular}{ccccc}
\hline
 Transition  		 & \mc{4}{c}{B(E2) [W.u.]  }	 \\
 $I_i^\pi \to I_j^\pi  $ & $\alpha_{30}$&$\alpha_{31}$&$\alpha_{32}$&$\alpha_{33}$ \\
\hline
 $3^- \to 1^-	$ & 166 & 20     &  276   &	274			  \\
 $5^- \to 3^-   $ & 140 & 138    &  412   & 	406			  \\
 $7^- \to 5^-   $ & 238 & 236    &  476   & 	468			  \\
 $9^- \to 7^-   $ & 308 & 306    &  512   & 	504			  \\
\hline
\hline
   		 & \mc{4}{c}{B(E1) [W.u.]  }	 \\
                 & $\alpha_{30}$&$\alpha_{31}$&$\alpha_{32}$&$\alpha_{33}$ \\
\hline          
 $3^- \to 2^+   $ & 6.0$\cdot10^{-3}$ & 3.8$\cdot10^{-3}$    &  2.6$\cdot10^{-3}$   & 1.2$\cdot10^{-4}$	  \\
 $5^- \to 4^+   $ & 2.6$\cdot10^{-3}$ & 1.4$\cdot10^{-3}$    &  3.0$\cdot10^{-3}$   & 2.0$\cdot10^{-4}$	  \\
 $7^- \to 6^+   $ & 1.3$\cdot10^{-3}$ & 7.2$\cdot10^{-4}$    &  3.0$\cdot10^{-3}$   & 1.4$\cdot10^{-4}$	  \\
 $9^- \to 8^+   $ & 8.2$\cdot10^{-4}$ & 4.4$\cdot10^{-4}$    &  3.2$\cdot10^{-3}$   & 2.2$\cdot10^{-4}$	  \\
\hline 
\end{tabular}
\end{table}

Nevertheless, the overall tendency of the $B(E1)/B(E2)$ ratio of preselected sequences of states called theoretical bands as function of spin, shown in parenthesis in Fig.~\ref{fig.07}, are directly extracted from the calculations.
First, for the negative-parity bands based on 1-phonon $\alpha_{3\mu}$ excitations, the lowering of the $B(E1)/B(E2)$ branching ratio with lowering spin, as discovered in
the experiment, is visible in bands constructed on $\alpha_{30}$ and $\alpha_{31}$ modes but in the latter, this ratio is about one order of magnitude 
higher than the experimental one of Ref.\cite{Har17}. As also seen, in the so called ''tetrahedral'' and
non-axial $\alpha_{33}$ bands this quantity is almost independent on spin. These facts may suggest that the experimental band, named in the article \cite{Har17} ''band 2'' 
may be most likely of axial octupole ($\alpha_{30}$) character. Note also that the non-axial $\alpha_{31}$ band efficiently competes with the latter.

The hypothesis that the vanishing of the intra-band $B(E2)$ transitions below $J=7^-$ state in studied negative-parity ''band 2'' of $^{156}$Dy nucleus can be provoked
 by presence of the ''tetrahedral'' symmetry is apparently not supported by this model.
The $B(E2)$ probabilities in the proposed ''tetrahedral'' band are even larger than in the axial-octupole band. 

What the dipole $E1$ transitions is concerned, in the axial $\alpha_{30}$ and non-axial $\alpha_{31}$ octupole bands the $B(E1)$ reduced transition probabilities 
grow monotonically by an order of magnitude with spin lowering from $9^-$ to $1^-$ while in the ''tetrahedral'' band-are almost unchanged.
Nevertheless, in all proposed model bands the magnitude of $B(E1)$ vary between $10^{-4}$ W.u. to $10^{-3}$W.u.
Since all the results are obtained within a pure collective approach, the $E1$ transition operator is constructed of the tensor couplings of 
the quadrupole and octupole ($\alpha_{2\mu}\otimes \alpha_{3\nu}$) modes, treated as the so called {\it second order contributions}, for details see Ref.\cite{myAPP2017}. 
To be detailed, the {\it first order contributions} to $E1$ transition operator would be proportional to $\alpha_{10}$, $\alpha_{1\pm 1}$ independent dipole deformations which are not considered in this work.
The presence of $\alpha_{3\nu}$ variables introduce a shift of the center of mass of nuclear surface with respect to the beginning of the coordinate system,
a spurious effect which should certainly be eliminated from the transition dipole operator. To cure this drawback 
we determine the so called {\it induced dipole deformations} as functions of independent variables $\alpha_{2\mu}$ and 
$\alpha_{3\nu}$ which inserted into the expansion (\ref{eqn.14}) translate the center of mass back to the beginning of the coordinate system. 
Here we profit of the approximate property of the $\alpha_{1}-$type deformations known to be responsible for the center-of-mass motion. As mentioned in \cite{myAPP2017} and references therein, such a shift is always accompanied by a modification of the surface shape. The stronger deformation $\alpha_1$, the larger change of the nuclear body is obtained.
Finally, let us notice that defined in such a way operator does not, by construction, take into account the microscopic effect of charge-density variation with the surface curvature, known as the polarization effects. 
We are convinced that for the discussed in this work low-lying collective configurations built in the ground state well and characterized by fairly compact shapes, this kind of effect is supposed in the first approximation to be negligible. For contrast, as concluded in \cite{myAPP2017}, effects related with center-of-mass shift can change the $B(E1)$
probabilities up to $\approx 40\%$.

The analysis of the theoretical ground-state and the negative-parity model bands reveals their tendency to be slightly ''squeezed compared to the experimental ones. 
This clearly is an indication that some ''fine tunning'' of the coupling constants of the rotational Hamiltonian, here obtained on the basis of the cranking moments of inertia
and Eq.~(\ref{coupl-const}) is needed. Please note that this constants are determined in the ground state point, and assumed to be constant during the vibrational motion. 
 In other words, a mechanism of vibration-rotation coupling through the deformation dependent moments of inertia is at this stage neglected.
On the other hand, it is interesting and remarkable that the relative energies of the octupole states with respect to the states of the equilibrium band are reproduced in a satisfactory way within some 0.1 MeV what may indicate on a reasonable predictive ability of the model.

\section{Summary}
\label{Section.04}

 Discussed model allow to construct the positive and negative-parity collective states based on $\alpha_{2\mu}$ and $\alpha_{3\nu}$ degrees of freedom.

Every vibrational-rotational state characterized by a given type of excitation and the number of excited phonons can occur in $J+1$ configurations described by
specific combinations of $K-$numbers which ensure the state to be symmetrizing with respect to the 
symmetrization $\bar{D}_{4y}$ group. Usually half of these states are lineary dependent, thus are not taken into account. Excitations in $\alpha_{30}$, $\alpha_{31}$ and $\alpha_{32}$ are close each other
 within $150$~keV.

 Presented results are preliminary and rather qualitative. The model does not reproduce perfectly the absolute values of $B(E\lambda)$, however the overall tendency of the $B(E1)/B(E2)$ ratio as function of spin can be directly extracted.

For the negative-parity bands based on one-phonon excitations in $\alpha_{3\mu}$ mode, the lowering of the $B(E1)/B(E2)$ 
with lowering spin, as discovered in
the experiment, is seen in the case of bands based on $\alpha_{30}$ and $\alpha_{31}$ modes. 
The so called ''tetrahedral'' and
non-axial $\alpha_{33}$ modes give this ratio almost independent on spin thus they are in contradiction the experimental observations. In $^{156}$Dy there are several measured collective bands but our interest is focused on ''band 2'', which structure has been particularly investigated in Ref.~\cite{Har17}.

The interesting point is the fact that
the K quantum number is not conserved within a band. Each excited rotational state is constructed as the superposition of contributions with different K values. This is the consequence of the symmetrization. Hence the amplitude of various K contributions may depend on the model components.

\end{document}